\documentclass[%
 reprint,
superscriptaddress,
 amsmath,amssymb,
 aps,
prb,
]{revtex4-1}

\usepackage{graphicx}
\usepackage{dcolumn}
\usepackage{bm}
\usepackage[utf8]{inputenc}
\usepackage{color}

\begin{document}


\title{Dynamic nuclear polarization at high Landau levels in a quantum point contact}
\author{M.~H.~Fauzi}
\email{fauzi@m.tohoku.ac.jp}
\affiliation{Center for Spintronics Research Network, Tohoku University, Sendai 980-8577, Japan}

\author{A. Noorhidayati}
\affiliation{Department of Physics, Tohoku University, Sendai 980-8578, Japan}

\author{M. F. Sahdan}
\altaffiliation[Present address: ]{Department of Physics, National University of Singapore, Singapore 117551, Singapore}
\affiliation{Department of Physics, Tohoku University, Sendai 980-8578, Japan}

\author{K. Sato}
\affiliation{Department of Physics, Tohoku University, Sendai 980-8578, Japan}

\author{K. Nagase}
\affiliation{Department of Physics, Tohoku University, Sendai 980-8578, Japan}

\author{Y. Hirayama}
\affiliation{Center for Spintronics Research Network, Tohoku University, Sendai 980-8577, Japan}
\affiliation{Department of Physics, Tohoku University, Sendai 980-8578, Japan}
\date{\today}

\begin{abstract}
We demonstrate a way to polarize and detect nuclear spin in a gate-defined quantum point contact operating at high Landau levels. Resistively-detected Nuclear Magnetic Resonance (RDNMR) can be achieved up to the $5$th Landau level and at a magnetic field lower than $1$ T. We are able to retain the RDNMR signals in a condition where the spin degeneracy of the first 1D subband is still preserved. Furthermore, the effects of orbital motion on the first 1D subband can be made smaller than those due to electrostatic confinement. This developed RDNMR technique is a promising means to study electronic states in a quantum point contact near zero magnetic field.

\begin{description}
\item[PACS numbers]
\end{description}
\end{abstract}

\pacs{Valid PACS appear here}
\maketitle


In a quantum Hall setting, generation of nuclear spin polarization and NMR detection are mostly carried out at the lowest Landau level (LL). Examples include quantum Hall edge channels\cite{Wald, Dixon, Machida01, Machida02, Corcoles, Keane, Chida, Fauzi}, breakdown of integer and fractional quantum Hall effect\cite{Kawamura07, Kou, Kawamura11, MStern, Tomimatsu, Miyamoto, Hashimoto16}, spin transitions of integer and fractional quantum Hall effect\cite{Kronmuller, Smet, Smet02, Hashimoto, Stern, Yusa, Kumada, FauziAPL, Friess, Tiemann, Rhone, Akiba, Moore}. A few Tesla magnetic field is typically required to reach the lowest LL.  Here, we advance the existing body of works by demonstrating local generation and detection of nuclear spin polarization in a quantum point contact at half-integer quantum Hall effect operating at up to the 5$^{\rm{th}}$ LL. This way we manage to push the lowest field down to less than 1 Tesla without relying on spin injection from ferromagnetic contacts \cite{Strand, Salis, Chan, Uemura}. 

One important application of this RDNMR techniques is to probe many body electronic states in the lowest 1D subband such as the $0.7 \times 2e^2/h$ anomalous conductance \cite{Thomas_1, Appleyard, Kristensen, Nuttinck, Cronenwett, Reilly_1, DiCarlo, Yoon, Sfigakis, Sarkozy, Ren, Mico, BrunPRL, Burke, FBauer, Iqbal, Kawamura15, Brun}, whose microscopic origin is still under active discussion. Early NMR measurement in the quantum point contact by Kawamura et al. \cite{Kawamura15} reveals that the $0.7$ anomaly does not arise from a bound state formation. The measurement itself is carried out in an in plane magnetic field of $4.5$ T to polarize the nuclei in a quantum point contact, but at the expense of almost fully lifting the spin degeneracy of the first 1D subband. It is not obvious whether the conclusion remains the same in the case where the spin is degenerate. In particular, a scanning gate microscope measurement by Brun et al. \cite{Brun, BrunPRL} observing an interference pattern of electron waves between a tip and a quantum point contact at zero magnetic field agrees with a single or multiple bound states formation scenario. Finding a way to perform NMR in a quantum point contact without lifting the spin degeneracy of the first 1D subband therefore is crucially important and hence the main focus of this paper.

Although in a quantum Hall setting the out of plane field exerts circular motion on electrons, we found that the effects on the 1D subband can be made smaller than those due to electrostatic confinement. This makes it possible to preserve the electrons motion at the lowest subband in the quantum point contact so that the motion is as close as possible to that without a magnetic field.

\begin{figure}[t]
\begin{center}    
\centering
\includegraphics[width=\linewidth]{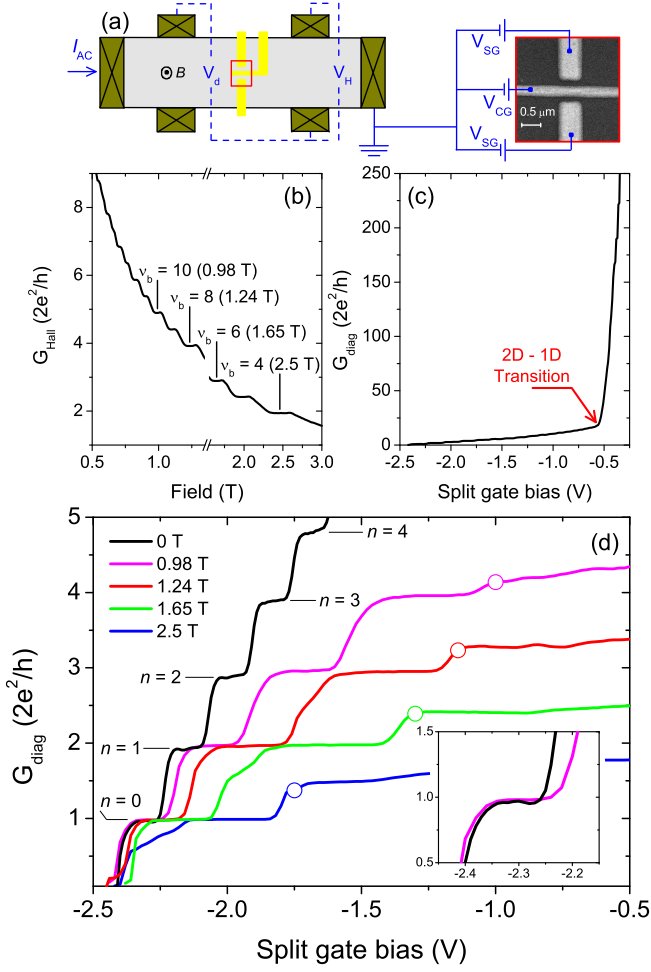}
\end{center}
\caption{(a) Device schematic and measurement setup. (b) Hall conductance $G_{\rm{Hall}}$ as a function of a perpendicular magnetic field in which several bulk even filling factors $\nu_{\rm{b}}$ are indicated. The sheet 2DEG electron density is $n = 2.36 \times 10^{15}$ m$^{-2}$. The constriction is open during the conductance recording. (c) zero magnetic field $G_{\rm{diag}}$ conductance profile as a function of split gate bias. 2D-1D transition is indicated by the arrow at approximately $V_{\rm{SG}} \approx -0.5$ V. (d) Diagonal conductance $G_{\rm{diag}}$ as a function of split gate bias voltage measured at several perpendicular magnetic fields. The open circles indicate the operating split gate bias at which RDNMR presented in the subsequent figure is measured. The corresponding subband indices $n$ are also indicated in the panel. The inset shows a blow-up of the first conductance profile for zero and 0.98 T fields. }
\label{Fig01} 
\end{figure}

\begin{figure}[t]
\begin{center}    
\centering
\includegraphics[width=\linewidth]{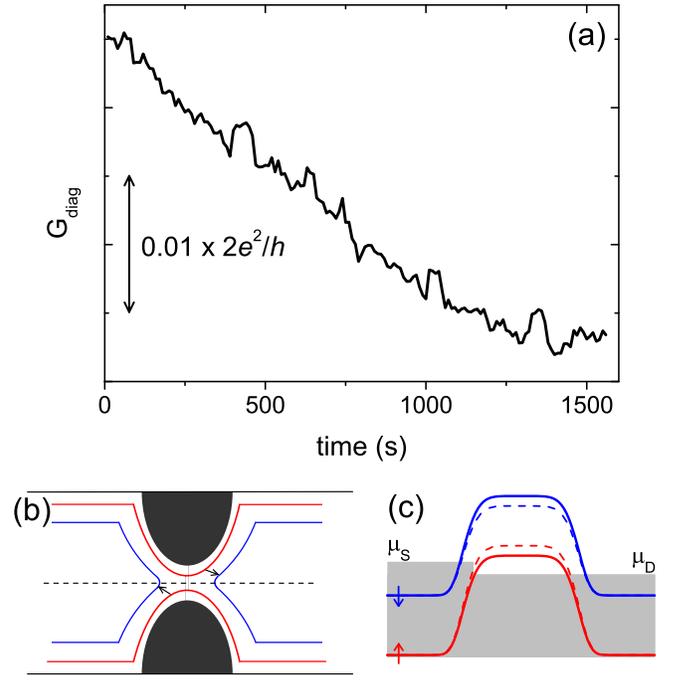}
\end{center}
\caption{(a) A typical time trace of conductance during current-induced dynamic nuclear polarization. (b) A schematic of an edge channel setup to create local nuclear spin polarization in a quantum point contact. The Red and blue lines represent up and down electron spin channels. For the sake of clarity, we only picture a pair of up and down spin channel that can belong to any Landau level. The black arrows represent the likely hyperfine-mediated spin-flip scattering process. The scattering event creates positive nuclear polarization (parallel to the external field) in and around the quantum point contact. (c) The corresponding potential barrier for up and down electron spin along the black dashed line in panel (b) without (solid line) and with (dashed line) positive nuclear polarization build-up.}
\label{FigEdge} 
\end{figure}

\begin{figure*}[t]
\begin{center}    
\centering
\includegraphics[width=\linewidth]{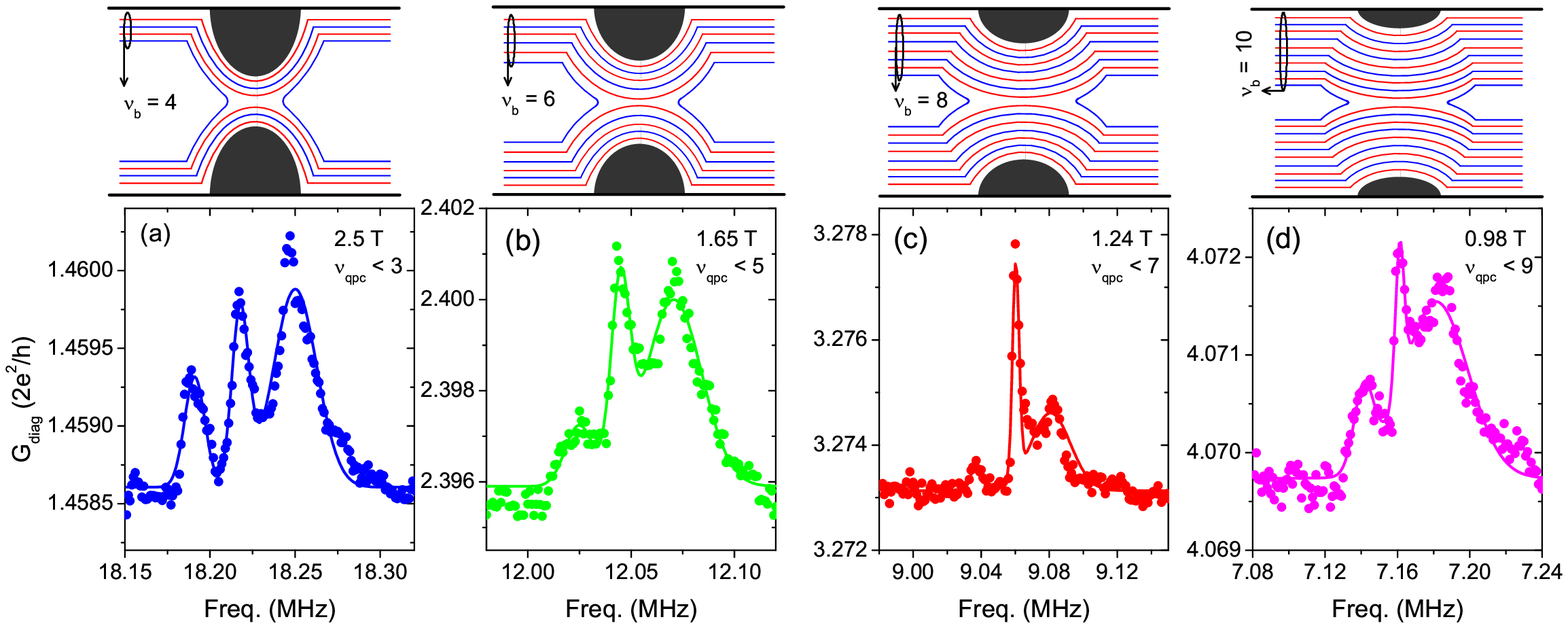}
\end{center}
\caption{(a)-(d) $^{75}$As RDNMR spectra taken at perpendicular fields of $2.5$ T ($\nu_{\rm{b}} = 4$, $\nu_{\rm{qpc}} < 3$), $1.65$ T ($\nu_{\rm{b}} = 6$, $\nu_{\rm{qpc}} < 5$), $1.24$ T ($\nu_{\rm{b}} = 8$, $\nu_{\rm{qpc}} < 7$), and $0.98$ T ($\nu_{\rm{b}} = 10$, $\nu_{\rm{qpc}} < 9$). The RF is swept with increasing frequency at a scan rate of $0.1$ kHz/s and RF power of $-30$ dBm delivered to the top of the cryostat. $G_{\rm{Hall}}$ and $G_{\rm{diag}}$ are simultaneously recorded with five subsequent traces averaged to increase signal-to-noise ratio. The appearance of three fold splitting noted in the $G_{\rm{diag}}$ is due to electric quadrupole interaction. The solid line is Gaussian fit to the data.}
\label{Fig03} 
\end{figure*}

The layout and dimensions of the present devices displayed in Fig. \ref{Fig01}(a) are similar to those used by Fauzi et al.\cite{Fauzi} with three independent Schottky gates to define the quantum point contact. Throughout the measurement process, we always equally biased a pair of split gates ($V_{\rm{SG}}$) and set the bias voltage zero to the center gate ($V_{\rm{CG}}$). Hall ($V_{\rm{H}}$) and diagonal voltage ($V_{\rm{d}}$) were simultaneously measured with phase-locked lock-in amplifiers at a frequency of $173$ Hz. We drove an AC current of $1$ nA for transport or $10$ nA for current-induced dynamic nuclear polarization (DNP) and NMR detection. The device was wrapped with coils to apply an oscillating RF magnetic field in the plane of 2DEG. The power delivered to the top of the cryostat was $-30$ dBm. It is worthwhile to mention that the delivered power has to be experimentally established to get an optimum signal. For instance, too high or too low RF power would reduce the NMR signal. We put the sample inside a dry dilution refrigerator with an electron temperature of about $100$ mK. The temperature was estimated by comparing the temperature dependence of an odd filling factor with the RF frequency slightly off-resonance.

The generic non-interacting energy spectrum of a quantum point contact in the presence of perpendicular magnetic field $B$ for any given subband index $n$ and an electron spin $S_z$ is\cite{Dixon}
\begin{equation} \label{eq:1}
E(n,S_{z}) = (n + \frac{1}{2})\hbar\omega + g_n^{*}\mu_{B}BS_{z} + AI_{z}S_{z}
\end{equation}
The first term describes the magneto-electric subband mixing with $\omega = \sqrt{\omega^2_y + \omega^2_c}$. $\omega_y$ and $\omega_c = eB/m^{*}$ are lateral electrostatic confinement and cyclotron frequency, respectively. The second term describes electron spin Zeeman energy with effective Lande-factor $g_n^{*}$, whose magnitude is subband-index dependent. The third term describes the contact hyperfine interaction between an electron and a nuclear spin. For a GaAs semiconductor, the conduction band has an s-wave like character resulting in a relatively strong contact hyperfine coupling of about $A = 85$ $\mu$eV\cite{Coish}. GaAs has three active nuclear isotopes with spin-3/2 namely $^{75}$As, $^{69}$Ga, and $^{71}$Ga.

Forcing electron spin to undergo spin flip scattering induces dynamic nuclear polarization (DNP) through hyperfine-mediated spin flip-flop interaction. The Hamiltonian is usually expressed as
\begin{equation}
    H_{hf} = AI \cdot S = \frac{A}{2}(I_{-}S_{+} + I_{+}S_{-}) + AI_{z}S_{z}
\end{equation}
The first term describes dynamic processes that make DNP possible. The last term describes static interaction between electrons and nuclear spins that make electric detection of NMR possible\cite{Hirayama}.

We will first present the basic transport characteristics of our quantum point contact device to pre-determine a ``hotspot'' to perform local DNP and RDNMR measurements. The bulk electron density was equal to $n = 2.36 \times 10^{11}$ cm$^{-2}$ with mobility $\mu = 2.31 \times 10^6$ cm$^2$/Vs. Fig. \ref{Fig01}(b) displays bulk 2DEG Hall conductance $G_{\rm{Hall}}$ as a function of a perpendicular magnetic field measured when the constriction was not formed by setting $V_{\rm{SG}} = 0$ and $V_{\rm{CG}} = 0$. The field was swept from $0.5$ to $3.0$ T. Notable bulk integer filling factors $\nu_{\rm{b}}$ are indicated in the panel.

The conductance profile measured at zero magnetic field revealed that the electron density beneath a pair of split gates was fully depleted at approximately $V_{\rm{SG}} \approx -0.5$ V, marked by a sharp decrease in the conductance as clearly displayed in Fig. \ref{Fig01}(c). The transport then switched from 2D to 1D regime when $V_{\rm{SG}} \le -0.5$ V. We ensured that all RDNMR measurements were carried out in a regime where the point contact was formed.

Fig. \ref{Fig01}(d) displays diagonal conductance vs split gate bias voltage measured at selected magnetic field values $B = 2.5, 1.65, 1.24$, and $0.98$ T, determined from the Hall conductance measurement results displayed in Fig. \ref{Fig01}(b). The corresponding bulk filling factors are $\nu_{\rm{b}} = 4, 6, 8$, and $10$, respectively. If we compare the conductance traces for each selected field, one can immediately see that the length of each integer plateau reduces as the subband index is lowered. A similar thing happens for the half-integer plateau. This can be understood if we consider the fact that as the subband index is lowered, the channel width gets narrower, so that the plateau length is mostly determined by the electrostatic confinement potential.

The inset to Fig. \ref{Fig01}(d) highlights the first integer conductance plateau for 0 and $0.98$ T magnetic fields. As expected, the length of the plateau is slightly extended as expected from magneto-electric subband mixing\cite{Wees88}. The wiggles seen on the zero conductance line occur due to Fabry-Perot resonances at the point contact with a flat-top potential barrier \cite{Heyder, Maeda}. However they disappear when applying a finite perpendicular field due to skipping orbit suppressing backscattering events\cite{Wees91}. Let us consider the cyclotron radius at $B = 0.98$ T is about 82 nm, which is about three times larger than the channel width of the first mode of about $25$ nm (roughly estimated from the Fermi wavelength of about $50$ nm). In this case, one can expect that the electrons motion is fairly close to that without a magnetic field. Namely, it proceeds mainly through the center of the constriction although the electron motion outside the constriction is dictated by quantum Hall physics.

We then proceeded to locally generate and detect nuclear spin polarization at high LL. For each selected magnetic field, we fixed the operating split gate bias to the left edge of the odd filling factors as displayed in Fig. \ref{Fig01}(d) to maximize the conductance sensitivity to a change in the Zeeman energy of electrons by the nuclear spin polarization\cite{Cooper}. An ac current of $10$ nA was applied to induce a DNP process as long as $1500$ seconds before sweeping the RF magnetic field through the coil. A typical time trace of conductance due to current induced DNP is displayed in Fig. \ref{FigEdge}(a). The forward spin-flip scattering events schematically shown in Fig. \ref{FigEdge}(b) produce positive nuclear polarization in the constriction, parallel to the external field. Although we only picture a pair of up and down electron spin channels, the pair and the associated spin-flip scattering process can belong to any Landau level. How does the conductance of the point contact change in the presence of positive nuclear spin polarization? Our picture indicates that the electron Zeeman splitting is reduced in the point contact so that the up spin electrons should see an increase in the barrier potential as schematically displayed in Fig. \ref{FigEdge}(c). This is the reason that, during current induced DNP, the conductance reduces with time. When the RF field hits the Larmor frequency of $^{75}$As, $^{69}$Ga, or $^{71}$Ga isotopes, destroying their polarization, the conductance increases (enhanced transmission)\cite{Fauzi}.



\begin{figure}[t]
\begin{center}    
\centering
\includegraphics[width=\linewidth]{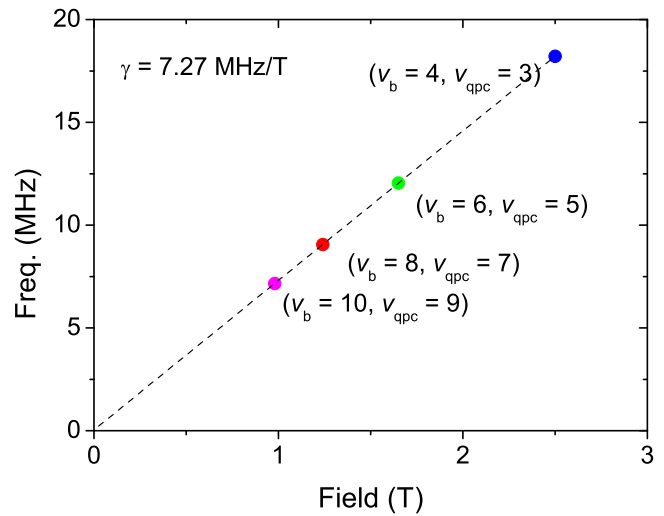}
\end{center}
\caption{Center transition As resonance frequency detected at different fields. The dashed line is the linear fit to the data extrapolated to zero field. The measured slope is about $7.27$ MHz/T.}
\label{S1} 
\end{figure}

The lower panels of Fig. \ref{Fig03} show $^{75}$As RDNMR spectra measured at the spin-split of the 2nd LL all the way up to the 5th LL. The corresponding edge state pictures are schematically displayed on the upper panels. We repeated the measurements five times and the results were averaged to increase the signal to noise ratio. For all the displayed spectra, we observed that the response substantially increases the conductance, which is consistent with our suggested picture displayed in Fig. \ref{FigEdge} (b)-(c). The three fold splitting due to strain-induced electric quadrupole interaction is clearly visible in all the spectra. The strain mainly originates from differential thermal contraction between the gate metal and GaAs semiconductor. Furthermore, the relative intensity of each peak varies depending on the operating field, which is likely to reflect the occupancy probability of each nuclear spin level after current-induced DNP \cite{Ota}. The measured $^{75}$As gyromagnetic ratio plotted in Fig. \ref{S1} is $\gamma = 7.27$ MHz/T, slightly smaller than the expected value of $\gamma = 7.29$ MHz/T. We attribute the slightly smaller gyromagnetic value of Arsenic nuclei to a combination of three different factors. The first is a finite Knight shift felt by the nuclei polarized in the point contact. The second is a small zero field offset in the superconducting magnet\cite{Kou, MStern}. The third is flux pinning in the superconducting wire making the magnetic field dependent on the ramping history of the superconducting solenoid.

It is interesting to note that the splitting reduces from about $30$ kHz measured at filling ($\nu_{\rm{b}} = 4$, $\nu_{\rm{qpc}} = 3$) displayed in Fig. \ref{Fig03}(a) to about $20$ kHz measured at filling ($\nu_{\rm{b}} = 10$, $\nu_{\rm{qpc}} = 9$) displayed in Fig. \ref{Fig03}(d). The electric quadrupole interaction itself does not depend on the magnetic field but on the strain field, provided that the field orientation with respect to the nuclear spin primary axis is maintained as is the case here. One plausible explanation for the quadrupole splitting variation detected here is the presence of spatial strain modulation imposed by the GaAs semiconductor and those three metal gates. The polarized nuclear spins could feel different strain fields around the point contact, but we will leave a detailed quantitative discussion of this for future publication.

\begin{figure}[t]
\begin{center}    
\centering
\includegraphics[width=\linewidth]{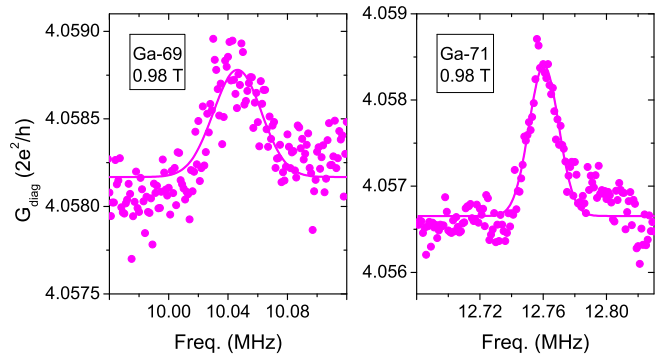}
\end{center}
\caption{$^{69}$Ga and $^{71}$Ga RDNMR spectra detected at the same condition as Fig. 3(d). The solid line is the Gaussian fit to the data with the corresponding linewidth of 38 kHz and 22 kHz, respectively.}
\label{S2} 
\end{figure}

We convinced ourselves that the signal at a field of $0.98$ T is truly an NMR signal since we were able to detect the signal coming from $^{69}$Ga and $^{71}$Ga nuclei as displayed in Fig. \ref{S2}. The spectra were detected under the same condition as in Fig. 3(d) (i.e. the 5th Landau level). The solid line in Fig. \ref{S2} is the Gaussian fit to the data with the corresponding linewidth of about $38$ kHz for $^{69}$Ga and $22$ kHz for $^{71}$Ga nuclei, respectively. Although the three-fold splitting is not resolved, the excess broadening is due to electric quadrupole interaction. Since the electric quadrupole moment $Q$ for $^{69}$Ga ($0.19 \times 10^{-28}$ m$^{2}$) is about 1.6 times bigger than $^{71}$Ga ($0.12 \times 10^{-28}$ m$^{2}$), so is the spectrum linewidth.

In summary, we have achieved generation and detection of nuclear spin polarization in quantum Hall setting at high LL of up to the 5th  LL without lifting the spin degeneracy of the first subband. Our developed NMR technique can potentially work at even lower perpendicular magnetic fields and/or elevated temperatures by employing a higher mobility device\cite{Rossler} and lower 2DEG density\cite{Kou}. For instance, a quantum point contact device used by A. Kou et al. \cite{Kou} has the 2DEG density three times lower than ours and the mobility more than five times higher than ours. This means that with such device, RDNMR detected at the $5^{\rm{th}}$ LL can be operated at a field as low as $0.3$ T and at elevated temperature. We believe that this would bring RDNMR technique to a new regime suitable for studying anomalous conductance in the lowest 1D subband. In the future, the presented technique will be applied to study electron spin dynamics by measuring nuclear spin relaxation rate ($T_{1}^{-1}$)\cite{Cooper} as well as static electron spin polarization in the point contact that can be deduced from RDNMR Knight shift\cite{Kawamura15}.

Beyond the anomalous conductance, the developed RDNMR technique we have described would be useful for probing inter-edge mode interaction. One example would be studying spin mode switching at the edge of a quantum Hall system recently suggested by Khanna et al \cite{Udit}.

We would like to thank K. Muraki of NTT Basic Research Laboratories for supplying high quality wafers for this study. We thank K. Hashimoto, B. Muralidharan, T. Aono, and M. Kawamura for helpful discussions. K.N. and Y.H. acknowledge support from Graduate Program in Spintronics, Tohoku University. Y.H. acknowledges financial support from KAKENHI Grants Nos. 15H05867, 15K217270, and 18H01811. MHF  acknowledges financial support from KAKENHI Grant No. 17H02728.

%

\end{document}